\documentclass[amsmath,showpacs,twocolumn,aps,prb]{revtex4}
\usepackage{bm}
\usepackage{graphics}
\usepackage{graphicx}
\begin{document}
\title{Plasmons in the presence of Tamm-Shockley states with Rashba splitting at noble metal surfaces}
\author{A.~M.~Farid}
\author{E.~G.~Mishchenko}
\affiliation{Department of Physics, University of Utah, Salt Lake
City, Utah 84112, USA}
\begin{abstract}

Au(111) or similar noble metal surfaces feature Tamm-Shockley
surface states that are known to possess considerable spin-orbit
splitting of the Rashba type of order $\Delta=0.1$ eV. When
interacting with an electromagnetic field such states are expected
to have resonances when the frequency of the field is near the
energy of the spin-orbit splitting $\Delta$. They originate from the
intersubband transitions between spin-split subbands and can be
observed in the frequency dependence of the surface impedance.
Plasmons in thin metal films are gapless and can be strongly
affected by these spin resonances, acquiring significant
modification of the  spectrum when it intersects the $\omega=\Delta$
line. Finally, an interesting demonstration of the  intersubband
resonances can be achieved when metal films are coated with ionic
dielectrics that have a frequency of longitudinal/transverse optical
phonons above/below $\Delta$. The dielectric function between the
two optical phonon frequencies is negative which forbids propagation
of conventional plasmon-polaritons. However, the presence of
spin-orbit-split surface states allows plasmon-polaritons to exist
in this otherwise forbidden range of frequencies.

\end{abstract}


\maketitle

\section{Introduction}
Nanoplasmonics is a novel field that emerged at the confluence of
optics and condensed matter physics \cite{HAA,SAM,BK}. Its ultimate
goal is the development of high resolution imaging methods by means
of plasmon-enhanced near-field optical measurements. Plasmons are
collective charge excitations of electron liquids that are induced by
external electric fields. In a {\it bulk} metal they are purely
longitudinal and have a gapped spectrum given by the well-known
Langmuir expression \cite{PN},
\begin{equation}
\Omega^2=\frac{4\pi e^2 N}{m}, \end{equation}
 in terms of bulk electron
density $N$ and mass $m$. At the interface between a metal and an
insulator (or vacuum) a distinct but closely related excitation can
propagate -- a surface plasmon, whose frequency is reduced
significantly, $\omega_s=\Omega/\sqrt{1+\kappa}$, by the dielectric
constant of the insulator $\kappa$ \cite{TS,HR}. In thin metallic
films such surface plasmons propagating near opposite interfaces
become hybridized and split into a symmetric mode (with oscillating
charges at the interfaces having the same sign) and anti-symmetric
mode (opposite sign). As a result the frequency of the
anti-symmetric mode increases while the symmetric mode becomes {\it
gapless}.
\begin{figure}
\includegraphics[width=6.5cm,angle=0]{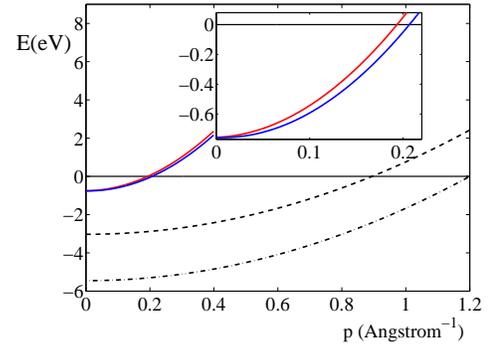}
\caption{(color online) Energy spectrum of Au(111) films relative to
the Fermi level (E=0): solid lines show the spectra of spin-orbit-split
Tamm-Shockley states, $E=p^2/2m\pm \alpha p/2$ vs.\ the in-plane momentum.
Dashed lines show the bulk electron spectrum $E=(p^2+p_z^2)/2m$ for $p_z=0; 0.2 \AA^{-1}$.} \label{fig.1}
\end{figure}

An electron liquid confined to {\it two dimensions} (2DEG),
created in semiconductor heterostructures, features a plasmon
spectrum remarkably different from its three-dimensional
analog. In particular, due to weaker screening the plasmon
spectrum of a single 2D layer is gapless \cite{Stern},
\begin{equation}
\label{2Dplasmon} \omega^2_k =\frac{2\pi e^2 n}{\kappa m^*} k,
\end{equation}
where $\kappa$ is a dielectric constant of a host semiconductor, $n$
is the planar density of the 2D electron system and $m^*$ is its
effective mass, typically significantly lower than the vacuum
electron mass $m_0$. Another celebrated property of a
two-dimensional electron system is the spin-orbit interaction. As
known from the relativistic Dirac equation, in the presence of {\it
any} potential $U({\bf r})$ the electron spin operator $\hat {\bf
s}$ is coupled to its momentum ${\bf p}$ via the Hamiltonian,
\begin{equation}
\label{hso} \hat H_{so} =-\lambda \nabla U \cdot (\hat {\bf s}\times
{\bf p}).
\end{equation}
Hereinafter we use units with $\hbar=1$.

 In a bulk crystal an
aperiodic part of the potential $U$ originates from impurities and
is relatively weak in a clean crystal. In a 2DEG, on the other hand,
it is {\it intrinsically} present as it comes from a confining
interface potential which creates the 2DEG \cite{BR}. Interestingly,
the value of the coupling constant $\lambda$ in a typical GaAs
semiconductor is six orders of magnitude stronger than in the vacuum
(and of opposite sign), where it is found from the Dirac equation to
be $\lambda=-1/2m_0^2 c^2 \approx -2\times 10^{-6}~ \AA^2$. Despite
this favorable fact, the spin-orbit splitting of a 2D electron
spectrum is rather weak \cite{NAT}: at the Fermi level, $\Delta \sim
0.1-1~ \text{meV}$, since the typical Fermi momenta are small, $\sim
0.01~\AA^{-1}$. As a result, the spin-orbit splitting leads to
relatively weak corrections \cite{MCE,XFW} to the 2D plasmon
spectrum, Eq.~(\ref{2Dplasmon}), and is difficult to observe
experimentally.

In the present paper we report on a novel phenomenon, which {\it
combines} the properties of bulk metals and two-dimensional electron
systems, and predict a resonant coupling of individual electron spin
degrees of freedom to a collective charge excitation originating
from the spin-orbit interaction.

Photoemission experiments performed on (111) surfaces of gold
\cite{LMJ,NRH,MHo,DP,HC} reveal the existence of $L$-gap surface
Tamm-Shockley \cite{Tamm,WS} states that constitute a
two-dimensional electron gas. Its properties are described by the
Fermi momentum $p_{F} \approx 0.2~\AA^{-1}$ and effective mass
$m^*=0.2m_0$ \cite{NRH}; for comparison, the Fermi momentum for bulk
electrons is $\approx 1.2~\AA^{-1}$. Of significant importance to us is
the observation that the spin-orbit splitting of the surface states
is unusually strong, which
 is measured to be $\Delta \approx 0.1~\text{eV}$ at the Fermi level. Recent experiments \cite{CRA} show that Bi/Ag(111) surface alloys can have a splitting which reaches $\Delta\approx 0.2~\text{eV}$.
 As we demonstrate below it is this strong spin-orbit interaction that
leads to coupling of electron spin to charge oscillations via
resonant electron transitions between spin-split subbands.
Fig.~1 shows the energy spectrum of both bulk and surface states for Au(111).

\subsection{Geometry of the system}

We will consider two geometries: a) a noble metal film of thickness
$d$ embedded into a insulator with a dielectric function  $\kappa
(\omega)$; b) a single metal interface with a vacuum. For the former
geometry we are interested in the plasmon-polariton spectrum, while
the geometry b) is discussed in relation to the observation of the
surface state resonances in the reflection coefficient (or
impedance) of electromagnetic radiation incident on the surface.

Both geometries require first that we analyze the dynamic response
of both surface and bulk electrons. The latter are conventionally
described by the Drude dielectric function
\begin{equation}
\label{eps3} \varepsilon(\omega) =-\frac{\Omega^2}{\omega
(\omega+i/\tau)},
\end{equation}
where the bulk plasma frequency in gold is $\Omega =1.3\times
10^{16} s^{-1}$. The electron momentum relaxation rate $1/\tau$ is
typically much smaller and depends on temperature (phonon
scattering) and crystal quality (impurities). Expression
Eq.~(\ref{eps3}) is the approximation that neglects vacuum
contribution and $d$-band electrons. However, the resonant phenomena
addressed in our paper occur at infrared frequencies, $\omega \sim
100~ \text{meV}$. At these frequencies the $s$-band contribution is
large and dominant, and  Eq.~(\ref{eps3}) to be a good
approximation.

The Hamiltonian of the surface states
\begin{equation}
\label{ham} \hat H=\frac{p^2}{2m^*} + \alpha (\hat{\bf s}\times {\bf
p})_z,~~~{\bf p}=(p_x,p_y),
\end{equation}
describes the in-plane dynamics of confined electrons, with the
effective spin-orbit coupling constant given by the average of
Eq.~(\ref{hso}) over the direction  perpendicular to the surface
($z$),
\begin{equation}
\label{alpha} \alpha = -\lambda \int\limits_{-\infty}^\infty dz
|\psi_{\bf p}(z)|^2 \frac{dU}{dz},
\end{equation}
where $\psi_{\bf p}$ is the wave function of a surface state. We
make an approximation (consistent with the photoemission data) that
$\alpha$ is momentum-independent. The numerical value of $\alpha$ is
related to the observed value of spin-orbit splitting according to
$\Delta=\alpha p_F$.

\section{Dynamic response of surface states}

Qualitatively, the interaction of the electromagnetic field (plasmon,
infrared beam reflected from the surface, etc.) with our system can
be understood as follows. An electric field along the $x$-direction
causes charge oscillations of {\it both} bulk and surface electrons
and an induced electric current $j_x$. It is the resonant
contribution to this current from the interband 2D transitions which
is of the most interest to us here. The result of the appearance of
the electric current $j_x$ is two-fold:

 i) According to Ampere's law,
a magnetic field along the $y$-direction appears that leads to the
interface Zeeman magnetization $M_y$.

ii) As seen from Eq.~(\ref{ham}) a drift along the
$x$-direction can be considered as effective ``Zeeman'' field
pointed along the $y$-direction. Consequentially, it induces
additional net electron spin polarization and, hence,
magnetization $M_y$ \cite{E}.

Subsequently, the surface magnetization and the accompanying ac
magnetization induce an additional ac electric field (along the
$x$-direction) via Faraday's law which acts upon both bulk and
surface electrons. It is crucial that all these effects are
resonantly {\it enhanced} for frequencies near the frequency of
inter-subband transitions, $\omega \sim \Delta$.

Let us now address the problem of dynamic response of 2D electron
states quantitatively. Most simply it can be done by means of a
kinetic equation. Due to the spin structure of the Hamiltonian
(\ref{ham}), the electron distribution function $\hat{f}_{\bf p}$ is
a $2\times2$ matrix in spin space. The corresponding equation
for $\hat{f}_{\bf p}$ in the presence of both ac electric and
magnetic fields has the form Ref.~\onlinecite{MH},
\begin{equation}
\label{kinetic}  \frac{\partial \hat{f}_{\bf p}}{\partial
t}+\frac{i}{2}\Delta_p ~ [\hat \eta_{\bf n} ,\hat{f}_{{\bf p}} ]=
\frac{i}{2}g\mu_B [\hat \sigma_y,\hat{f}_{{\bf p}}]H_y-eE_x
\frac{\partial \hat{f}_{{\bf p}}}{\partial p_x},
\end{equation}
where $\Delta_p=\alpha p$, $\hat \eta_{\bf n}=n_y\hat
\sigma_x-n_x\hat \sigma_y$,  with ${\bf n}={\bf p}/p$ being the
direction of electron momentum and $\hat {\bm \sigma}=2\hat{\bf s}$
the set of Pauli matrices; $\mu_B=|e|/2mc$ is the Bohr magneton and
$g$ is the gyromagnetic ratio of surface electrons.

A formal solution to the equation (\ref{kinetic}) in the frequency
domain reads (see Appendix A for more details)
\begin{eqnarray}
\label{anzats}  \hat f_{\bf p}= i\frac{(\Delta_p^2-2\omega^2)\hat
{\cal K}_{\bf p}+\Delta^2_p \hat{\eta}_{\bf n}  \hat {\cal K}_{\bf
p}\hat{\eta}_{\bf n}-\omega\Delta_p [\hat\eta_{\bf n},\hat{\cal
K}_{\bf p}]}{2\omega(\Delta_p^2-\omega^2)} ,
\end{eqnarray}
where ${\cal K}_{\bf p}$ denotes the right-hand side of
Eq.~(\ref{kinetic}). To the linear order in the fields $H_y$ and
$E_x$ it is sufficient to utilize
the equilibrium distribution function in the right-hand side of
Eq.~(\ref{kinetic}),
 $\hat f^{(0)}_{\bf p}=\frac{1}{2} \sum(1\pm\hat \eta_{\bf n})n_\pm$,
with $n_\pm =n_F (p^2/2m^* \pm \Delta_p -\mu)$ denoting the Fermi-Dirac functions for the two spin-split subbands.

Using Eq.~(\ref{anzats}) it is now starightforward to calculate
the surface magnetization ${\bf M}=-\frac{1}{2} g\mu_B \text{Tr}\int
\frac{d^2 p}{(2\pi)^2} \hat f_{\bf p} \hat {\bm \sigma}$, and
current density $j_x=e ~\text{Tr}\int \frac{d^2 p}{(2\pi)^2} \hat
f_{\bf p} (\frac{p_x}{m^*} -\frac{1}{2}\alpha \hat {\sigma}_y)$.
Note that the electron velocity  contains spin operator. After
momentum integration we obtain for the 2D  electric current and
magnetization,
\begin{eqnarray}
\label{response}
&& j_x=\sigma(\omega) E_x(0)-i\omega \beta (\omega) H_y(0),\nonumber\\
&& M_y=\beta(\omega) E_x(0) +\chi (\omega) H_y(0).
\end{eqnarray}
Here we explicitly emphasized that the fields are to be taken at the
interface ($z=0$) (see also the discussion in the next section.) Here
the electric conductivity of surface states  ($\Delta
=\Delta_{p_F}$),
\begin{equation}
\label{current} \sigma(\omega)= \frac{ie^2n}{\omega m^*}
+\frac{im^*}{8\pi\omega} \frac{e^2
\alpha^2\Delta^2}{\omega^2-\Delta^2},
\end{equation}
where $n=p_F^2/2\pi$ is the density of 2D electrons, consists of the
usual intrasubband Drude conductivity (first term) and the
contribution of resonant intersubband transitions (second term). The
dynamic magnetic susceptibility $\chi(\omega)$ also features a
similar resonant structure, which with the help of
Eq.~(\ref{anzats}) is found to be
\begin{equation}
\label{mag} \chi(\omega) =g^2 \mu_B^2\frac{m^*}{8\pi}
\frac{\Delta^2}{\omega^2-\Delta^2}.
\end{equation}
Finally, the cross-susceptibility has the form,
\begin{equation}
\label{spin} \beta(\omega) = \frac{iem^*}{8\pi}
\frac{g\mu_B\alpha\Delta^2}{\omega(\omega^2-\Delta^2)}.
\end{equation}
Note that the off-diagonal terms in Eqs.~(\ref{response}) are
related to each other, in agreement with the Onsager theorem.

In the presence of two surfaces, expressions similar to the above
Eqs.~(\ref{response}-\ref{spin}) hold for the other surface as well
(with the reversal of the sign of $\alpha$, which corresponds to the
reversal of the sign of $dU/dz$ in Eq.~(\ref{alpha}).

\section{Impedance of $\text{Au}$(111) surface}

The most straightforward way to observe resonances,
Eqs.~(\ref{current})-(\ref{spin}), associated with the intersubband
transitions is to measure the reflection of the electromagnetic wave
incident from vacuum on a surface of a bulk metal. Since typical
photon momenta $\Delta/c$ is considerably smaller than the
characteristic 2D electron momentum scales of the problem, $p_F$ and
$\Delta/v_F$, it is sufficient to consider a case of normal
incidence and reflection. Assuming the incident wave has frequency
$\omega$ and amplitude $E_0$, we can write the electric field as
\begin{equation}
\label{impedance_electric} E_x(z)=\left\{\begin{array}{ll} E_0 e^{-i
{\omega}z/c} +E_2 e^{i\omega z/c} ,~& 0<z,\\ E_1 \exp{(
{\omega}z\sqrt{-\varepsilon}/c)},~& z<0. \end{array} \right.
\end{equation}
\begin{figure}
\includegraphics[width=7.5cm,angle=0]{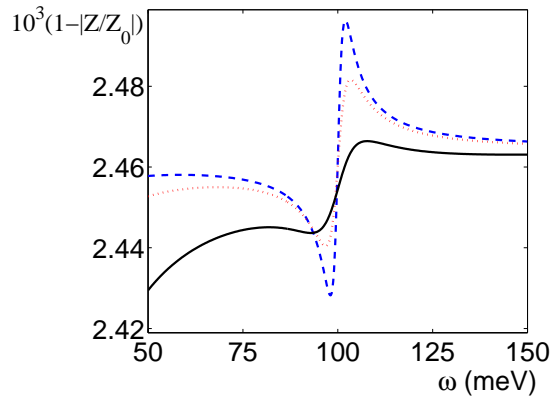}
\caption{(color online) Frequency dependence of the absolute value
of the surface Au(111) impedance $|Z(\omega)/Z_0(\omega)|$, as
related to the impedance $Z_0(\omega)
=1/\sqrt{-\varepsilon(\omega)}$ in the absence of Tamm-Shockley
states, for different values of the electron scattering rate:
$1/\tau=1\times 10^{12} \text{s}^{-1}$ (blue dashed line), $3\times
10^{12} \text{s}^{-1}$ (red dotted line), $1\times 10^{13}
\text{s}^{-1}$ (black solid line). The energy of spin-orbit-induced
intersubband resonance is $\Delta=100 \mbox{meV}=1.5\times 10^{14}
\text{s}^{-1}$.} \label{fig.3}
\end{figure}
Accordingly, the magnetic field $H_y$ is found from Maxwell's
equations. Inside the metal it reads $\partial H_y/\partial
z=i\omega\varepsilon E_x/c$; and similarly in the vacuum (where
$\varepsilon \to 1$). As a result,
\begin{equation}
\label{impedance_magnetic} H_y(z)=\left\{\begin{array}{ll} -E_0
e^{-i {\omega}z/c} +E_2 e^{i\omega z/c} ,~& 0<z,\\
-i\sqrt{-\varepsilon}E_1 \exp{({\omega}z\sqrt{-\varepsilon}/c)},~&
z<0.
\end{array} \right.
\end{equation}
Boundary conditions relate the discontinuities of the fields $E_x$
and $H_y$ to the surface current and magnetization given by
Eqs.~(\ref{response}),
\begin{eqnarray}
\label{boundary_conditions}
E_0+E_2-E_1=\frac{4\pi i \omega}{c} M_y,\\ -i\sqrt{-\varepsilon}E_1 +E_0-E_2=\frac{4\pi j_x}{c}.
\end{eqnarray}
The right-hand side of the boundary conditions
(\ref{boundary_conditions}) are not well defined as the surface
electric current and magnetization, Eqs.~(\ref{response}), are given
in terms of electric $E_x(0)$ and magnetic $H_y(0)$ fields right
{\it at the surface}. The latter, however, are discontinuous at
$z=0$. This means that the response of surface electrons has to be
in principle found from a solution of a  three-dimensional problem
which accounts for the $z$-dependence of the electron density
distribution, dynamics of surface states, as well as Maxwell's
equations in a self-consistent way. Such analysis is well beyond the
scope of the present paper and should be a subject of a separate
work. In this paper we adopt a phenomenological approach and
approximate that the 2D electron density is constant across the
interface so that that the effective surface electric and magnetic
fields in Eqs.~(\ref{response}) are given by the corresponding mean
arithmetic values of the magnitude of the fields immediately above
and below the surface,
\begin{equation}
\label{effective}
E_x(0)=\frac{E_0+E_1+E_2}{2},~~~H_y(0)=\frac{E_2-E_0-i\sqrt{\varepsilon}E_1}{2}.
\end{equation}
In Appendix B we show how these boundary conditions follow from the
approximation of 2D electron density constant across the interface.
Equations (\ref{boundary_conditions}) together with (\ref{response})
and (\ref{effective})  now give a system of two coupled linear
equations for $E_1$ and $E_2$. The experimentally measurable
quantity, surface impedance $Z(\omega)$, is given by the ratio of
the electric and magnetic fields at $z=0+$,
\begin{equation}
Z(\omega)=-\frac{E_x(0+)}{H_y(0+)}=\frac{E_0+E_2}{E_0-E_2}.
\end{equation}
After simple calculation we obtain
\begin{equation}
Z(\omega)=\frac{1+\left(\frac{2\pi\omega }{c}\beta\right)^2+
\frac{4\pi\omega}{c}\chi\left(\sqrt{-\varepsilon}-\frac{i\pi\sigma}{c}
\right)}{i\sqrt{-\varepsilon}\Bigl[1+\left(\frac{2\pi\omega
}{c}\beta\right)^2\Bigr]+\frac{4\pi\sigma}{c}\left(1+\frac{\pi\omega}{c}
\sqrt{-\varepsilon} \chi \right)}.
\end{equation}
Substituting now the values of $\sigma$, $\chi$ and $\beta$ found in
the preceding section we observe that the doubly-resonant terms,
$\propto \beta^2$, are exactly  canceled by the corresponding
contributions from the terms $\propto \sigma \chi$. Of the remaining
terms only the $\sigma$-term in the denominator is to be kept, as
the $\chi$-terms are small by virtue of $(v_F/c)^2(\Omega/\Delta)
\ll 1$,
\begin{equation}
\label{z}
Z(\omega)=\frac{1}{i\sqrt{-\varepsilon(\omega)}+{4\pi\sigma(\omega)}/{c}}.
\end{equation}
The resonant feature in Eq.~(\ref{z}) is rather narrow though sharp.
A finite electron scattering rate leads to its broadening.

\subsection{Electron scattering}

The scattering of surface electrons off phonons, impurities and
surface roughness can be be accounted for by a collision integral in
the kinetic equation (\ref{kinetic}). In a relaxation time
approximation this yields the substitution $\omega \to \omega
+i/\tau$ in the response functions (\ref{current})-(\ref{spin}). The
broadening of the bulk electron response is accounted for by the
imaginary part in the dielectric function (\ref{eps3}). For
estimates we will assume that the scattering rate $1/\tau$ is the
same for bulk and surface states. Fig.~\ref{fig.3} illustrates the
dependence of the absolute value of the impedance $|Z(\omega)|$
close to the resonant frequency for different values of the
scattering rate. Typical scattering rates in noble metals at room
temperature are $\sim 10^{14}~\text{s}^{-1}$. In order to achieve
the desired scattering rate of order $\sim 10^{12}~\text{s}^{-1}$
one should make measurements at low, $T <10~\text{K}$, temperatures.
Due to the screening from bulk electrons, which gives large values
for the refraction index $\sqrt{-\varepsilon} \approx
\Omega/\Delta$, the relative correction to the surface impedance is
small but within the capabilities of modern optical detection
methods.

An alternative setup for the observation of intersubband surface
resonances is the measurement of the transmission coefficient through a
thin metal film, rather than of a reflection (impedance) from a bulk
metal. The corresponding calculations can be performed in complete
analogy to the above analysis.

Finally, we note that the applicability of the linear response close
to the resonance requires that the first-order variation of the
electronic distribution function is small compared with the
equilibrium Fermi-Dirac distribution. Estimating the right-hand side
of Eq.~(\ref{anzats}) we obtain that close to the resonance the
following condition should be satisfied, $eE v_F/ \Delta \ll
\text{max} (|\omega-\Delta|, 1/\tau)$. This condition  has a simple
physical meaning -- the energy acquired over the spin-orbital
distance $v_F/ \Delta$ in the electric field should be smaller than
the detuning to the resonance or the scattering rate.

\section{Plasmon spectrum of a metal film}

\begin{figure}
\includegraphics[width=6.5cm,angle=0]{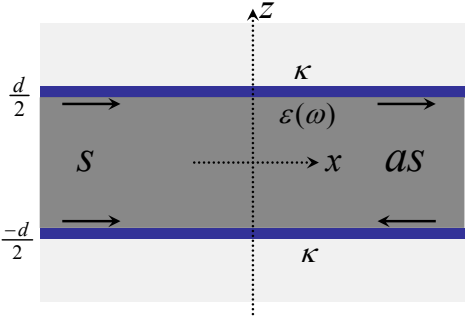}
 \caption{(color online)  Au(111) film of thickness $d$ embedded
 into an insulator with dielectric constant $\kappa$. Dark area
 contains bulk electrons described by $\varepsilon(\omega)$.
 Stripes near the film surfaces indicate 2DEG of Tamm-Shockley electrons.  Arrows show directions
 of $E_x$ for symmetric ($s$) and anti-symmetric ($as$) modes.} \label{fig.2}
\end{figure}

We now consider the role of intersubband resonances on the
properties of plasmons in thin Au(111) films, embedded into a
dielectric, see Fig.~\ref{fig.2}. The plasmon is propagating along the
$x$-direction. In general, it is accompanied by $E_x$, $E_z$
components of electric field as well as an $H_y$ component of magnetic
field. As illustrated in Fig.~\ref{fig.2} there are two modes in a
film, corresponding to symmetric and anti-symmetric alignment of the
electric field near the two surfaces.

Let us begin by estimating the relative importance of the two
resonant  contributions, arising from the surface electric current
and surface magnetization, cf. Eqs.~(\ref{response}). Integrating
Faraday's law, $\nabla \times {\bf E} =i\omega ({\bf H}+4\pi
{\bf M})/c$, across an interface,  we can obtain that the additional
electric field induced by the oscillating surface magnetization is
$\delta E_x^{(1)} \sim \omega M_y/c$. Similarly, according to
Ampere's law the additional contribution to the magnetic field due to
the surface current is $\delta H_y \sim \delta j_x/c$, where $\delta
j_x$ of interest is the second (resonant) term in
Eq.~(\ref{current}). Consequentially, this magnetic field yields an
additional electric field given by $\delta E_x^{(2)} \sim (\partial
H_y/\partial z)c/\omega\kappa \sim \delta j_x
\sqrt{k^2-\omega^2/c^2}/\omega\kappa$,  for a wave with the
wavevector ${\bf k}$ along the $x$-direction. Thus, the relative
magnitude of the two effects is
\begin{equation}
\label{est} \frac{\delta E_x^{(1)}}{\delta E_x^{(2)}} \sim
\frac{\omega^2\kappa M_y}{c\sqrt{k^2-\omega^2/c^2} j_x} \sim \frac{g
p_F \omega^2\kappa}{mc^2 \Delta \sqrt{k^2-\omega^2/c^2}},
\end{equation}
where in the last identity we made use of Eqs.~(\ref{response}) and
the response functions (\ref{current})-(\ref{spin}).  To obtain a
numerical estimate of the ratio in Eq.~(\ref{est}) it is sufficient
to utilize the plasmon spectrum for $\Delta=0$ (see
Eq.~(\ref{dispersion}) with the first two terms only). At low
frequencies, much less than the plasma frequency, $\omega \ll
\Omega$, we find, $\kappa/\sqrt{k^2-\omega^2/c^2} \approx
\frac{c\Omega}{\omega^2}\tanh(\frac{d\Omega}{2c})$. We thus obtain,
\begin{equation}
\label{estimate} \frac{\delta E_x^{(1)}}{\delta E_x^{(2)}}  \sim g
\frac{p_F}{mc}\frac{\Omega}{\Delta}~\tanh \left({\Omega
d}/{2c}\right).
\end{equation}
As seen from this expression, smaller values of spin-orbit splitting
$\Delta$ favors the role of magnetization. At present we do not know
the gyromagnetic ratio $g$ of surface electrons. For $g=2$ and
$\Omega=1.4\times 10^{12} \text{s}^{-1}$ we estimate $\sim g
\frac{p_F}{mc}\frac{\Omega}{\Delta} = 0.14$ for Au(111). The effect
of the electric current is thus considerably stronger than the effect of
magnetization. For thin films $d<c/\Omega$ the suppression of
$\delta E_x^{(1)}$ is even more noticeable. Yet, assuming that in
some other materials the situation can be different, in the
following we analyze both contributions.

 For a plasmon wave
propagating along the $x$ direction, the electric field ${\bf E} (z) e^{i
kx-i\omega t}$ obeys the wave equation, which inside the film
($|z|<d/2$) has the form,
\begin{eqnarray}
\label{maxwell}
\frac{d^2{\bf E}}{dz^2} -\left(k^2- \varepsilon \frac{\omega^2}{c^2}\right){\bf E}=0,
\end{eqnarray}
where the dielectric function $\varepsilon$ is still given by
Eq.~(\ref{eps3}). The corresponding wave equation outside the film
has the same form as Eq.~(\ref{maxwell}) with $\varepsilon \to
\kappa$. Since equation (\ref{maxwell}) is invariant under spatial
inversion transformation, $z\to -z$, its solutions  are either
symmetric or anti-symmetric under this inversion. The symmetric
solution can be written as
\begin{equation}
\label{Poisson_sol} E_x(z)=\left\{\begin{array}{ll}
E_1\frac{\cosh{(K_1 z)}}{\cosh{(K_1 d/2)}},~& |z|<d/2,\\  \\E_2
e^{-K_2|z|-d/2},~& d/2<|z|, \end{array} \right.
\end{equation}
where $K_1=\sqrt{k^2-\varepsilon \omega^2/c^2}$ and
$K_2=\sqrt{k^2-\kappa \omega^2/c^2}$. The anti-symmetric
solution can be written in the same manner only with the
change, $\cosh \to \sinh$.

The magnetic field is found similarly to how it is done in Sec.IV and
reads
\begin{equation}
\label{magn_sol} H_y(z)=\frac{i\omega}{c}\left\{\begin{array}{ll}
\frac{\varepsilon E_1}{K_1}\frac{\sinh{(K_1 z)}}{\cosh{(K_1
d/2)}},~& |z|<d/2,\\ \\-\text{sgn}{(z)}\frac{\kappa E_2}{K_2}
e^{-K_2|z|-d/2},~ & d/2<|z|.
\end{array} \right.
\end{equation}
The boundary conditions relate the discontinuity of the electric field
to the interface magnetization
\begin{equation}
\label{boundary_conditions_plasmon} E_2-E_1=\frac{4\pi i \omega}{c}
M_y, \end{equation}
and the discontinuity of the magnetic field to the
surface current,
\begin{equation}
\label{boundary_conditions_plasmon2} \ \frac{\varepsilon E_1
\tanh{(K_1 d/2)}}{K_1} +\frac{\kappa
E_2 }{K_2}=\frac{4\pi
j_x}{i\omega},
\end{equation}
cf.~Eqs.~(\ref{boundary_conditions}). As already pointed out in
Sec.~III, the right-hand sides of boundary conditions
(\ref{boundary_conditions_plasmon}-\ref{boundary_conditions_plasmon2})
are not well defined as the interface current and magnetization
(\ref{response}) have been calculated under the assumption that
the electric and magnetic fields are uniform across the
interface. Adopting the same approximation as utilized above,
cf.~Eq.~(\ref{effective}), we write
\begin{equation}
E_x(d/2)=\frac{1}{2}(E_1+E_2),
\end{equation}
and similarly for the magnetic field.  After straightforward
transformations we find the dispersion equation for symmetric
plasmons,
\begin{eqnarray}
\label{dispersion} \varepsilon(\omega)
\frac{\tanh{\frac{d}{2}K_1}}{K_1}&+&\frac{\kappa}{K_2}-\frac{4\pi \sigma (\omega)}{i\omega} \nonumber \\
&=&\frac{4\pi \omega^2}{c^2} \chi(\omega) \frac{\varepsilon(\omega)
\kappa \tanh{\Bigr[\frac{d}{2}K_1\Bigr]}} {K_1K_2 }.~~~~
\end{eqnarray}
The spectrum of anti-symmetric plasmons is determined by the
equation which is obtained from Eq.~(\ref{dispersion}) by the
substitution $\tanh \to \coth$. Qualitatively, the modification
of the spectrum of low-frequency anti-symmetric modes is
similar to that of symmetric modes though  quantitatively
smaller. Below we discuss symmetric plasmons only.

The first two terms in Eq.~(\ref{dispersion}) describe conventional
plasmons in a metallic film surrounded by a dielectric. For $d\to
\infty$ and $ck \gg \omega$ they yield a well-known surface  plasmon
dispersion relation $\varepsilon+\kappa=0$. The third term
originates from the {\it electric} response of Tamm-Shockley surface
electrons and contain an intrasubband Drude term as well as
an intersubband resonant contribution, cf. Eq.~(\ref{current}).
Finally, the last term describes the {\it magnetic} response of surface
states, which also features interband resonances at $\omega=\Delta$,
see Eq.~(\ref{mag}). As can be easily verified, the relative
magnitude of magnetic and electric 2D responses is indeed controlled
by the above parameter, Eq.~(\ref{estimate}).

At low frequencies, $\omega \ll \Omega$, and the dielectric function is
$\varepsilon \approx -\Omega^2/\omega^2$. Since wavelengths of
interest are small, $k \ll \Omega/c$, after simple transformations
(neglecting the last term in Eq.~(\ref{dispersion}),
 which is typically a good
approximation), we find a plasmon spectrum in the form,
\begin{equation}
k^2=\frac{\kappa \omega^2}{c^2} +\frac{\kappa^2
\omega^4}{\left(c\Omega \tanh{\frac{\Omega d}{2c}}+\frac{4\pi e^2
n}{m^*}+\frac{m^*e^2\Delta^4}{2p_F^2[(\omega+i/\tau)^2-\Delta^2]}\right)^2}
.
\end{equation}
Here we assumed that scattering rate $1/\tau$ is small compared with
frequency $\omega$ and retained it in the resonant intersubband term
only. For thin films, $d \ll c/\Omega$, which for Au means that
$d<10\text{nm}$, the electric field is almost uniform across the
film, so that oscillations of surface electron density simply add to
the oscillations of the charge density in the bulk of the film.

\begin{figure}
\includegraphics[width=8.0cm,angle=0]{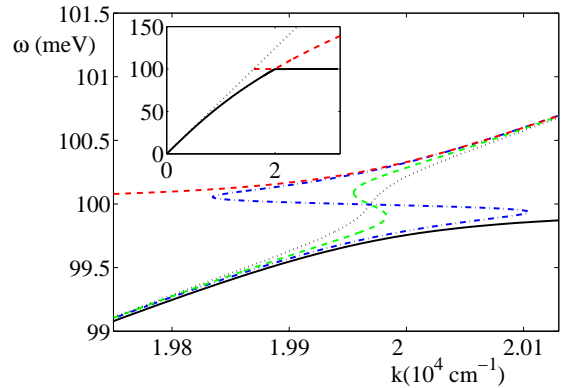}
 \caption{(color online)  The effect of the interband spin
resonance on the spectrum of symmetric plasmons in a thin Au(111) film,
$d=2~\text{nm}$, coated with a dielectric with $\kappa =10$. The
inset shows the full frequency range. The main plot shows
frequencies close to the resonant frequency $\Delta = 100~
\text{meV}$. The interaction with the surface electrons leads to
strong resonant modification of plasmon spectrum close to $\omega
=\Delta$. The plasmon spectrum is shown for different values of the
scattering rate: $1/\tau=1\times 10^{11}~\text{s}^{-1}$ (blue
dashed-dotted line), $3\times10^{11}~\text{s}^{-1}$ (green dashed
line) and $5\times10^{11}~\text{s}^{-1}$ (dotted black line). Red
dashed and black solid lines (shown also in the inset) represent the
asymptotes when $1/\tau \to 0$. The plasmon group velocity is varied
greatly in this frequency range and can change sign provided that
scattering rate is small enough. Note that the plasmon spectrum is
bounded by the photon line $~\omega=ck/\sqrt \kappa$ from the left.
} \label{fig4}
\end{figure}

Fig.~\ref{fig4} illustrates the effect of the interband spin
resonance on the plasmon spectrum of thin Au(111) films. Near the
point where this spectrum intersects the  spin resonance frequency,
$\omega =\Delta$, there is a significant decrease in the plasmon
group velocity, $\partial \omega/\partial k$, and even a narrow
region of {\it negative} group velocity. In addition, at
$\omega=\Delta$ the plasmon {\it phase} velocity is increased
significantly and (in the absence of electron scattering) reaches
the value of $c/\sqrt{\kappa}$, characteristic of photon propagation
in the dielectric. The corresponding increase of the wavelength
$1/k$ due to the resonant interaction with surface electrons can be
quite significant, as seen from Fig.~\ref{fig4}. A finite scattering
rate $1/\tau$ makes these features less {\it sharp}, but {\it
broadens} them, which could in fact help their observation. However,
the scattering rate should be made rather small, which emphasize the
use of low temperatures.

\section{Plasmon spectrum of a metal film embedded in an ionic dielectric}

 Another interesting manifestation of the surface states can be achieved if
the material surrounding the metal film is an ionic crystal. (Here
we assume that the presence of a dielectric, while possibly changing
the value of $\Delta$, does not suppress it significantly.) As
well-known, the  dielectric function of an ionic crystal \cite{AM},
\begin{equation}
\kappa(\omega)=\kappa_\infty\frac{\omega^2-\omega_L^2}{\omega^2-\omega_T^2},
\end{equation}
has a pole at the frequency of the transverse optical phonon $\omega_T$
and  a zero at the frequency of the longitudinal optical phonon
$\omega_L$. Since the dielectric function is negative within the
frequency range, $\omega_T<\omega<\omega_L$, the plasmon (or, more
accurately, plasmon-polariton) excitations normally cannot propagate
in this range of frequencies. This can be verified from
Eq.~(\ref{dispersion}) with $\Delta=0$, as the first two terms in
this equation are {\it negative}.

However, the finite spin-orbit splitting can change this  situation
dramatically if the splitting energy $\Delta$ falls within the
forbidden frequency band, i.e.  $\omega_T<\Delta<\omega_L$. For a
Au(111) film with $\Delta\approx 100~ \text{meV}$ such a situation
can be realized, in particular, for LiH, $\kappa_\infty =3.6$,
$\omega_T= 72~ \text{meV}$, $\omega_L= 138~ \text{meV}$ or SiC,
$\kappa_\infty =6.7$, $\omega_T= 98 ~\text{meV}$, $\omega_L= 117
~\text{meV}$, data from Ref.~\onlinecite{K}. Since the resonant term
in Eq.~(\ref{current}) {\it changes sign} below $\omega =\Delta$,
plasmon-polariton modes can propagate in the bandgap with a
frequency close to the frequency of intersubband transitions.

For thin films $d \ll c/\Omega$ the bandgap resonance is at
\begin{equation}
\label{last} \omega=\Delta -\frac{m^*e^2 \Delta^3}{2dp_F^2\Omega^2}.
\end{equation}
The impedance of the coated film can be obtained  similarly to the
calculations of the preceding Sections. We assume here that the
thickness of the insulator coating $D$ is smaller than the decay
length of electromagnetic field in the insulator, $D \ll c/\Delta <
2~ \mu m$. The impedance is then found to be
\begin{equation}
\label{prelast} Z(\omega) =\frac{2\Delta c}{id\Omega ^2}
\left(1-\frac{\frac{m^*e^2
\Delta^4}{dp_F^2\Omega^2}}{(\omega+i/\tau)^2-\Delta^2+\frac{m^*e^2
\Delta^4}{dp_F^2\Omega^2}}  \right).
\end{equation}
We obtain that the width of the resonance is given by the scattering
rate and its relative amplitude is proportional to the ratio of the
second term in Eq.~(\ref{prelast}) and the scattering rate. For the
metal film of thickness $d=10~\text{nm}$ and scattering rate
$1/\tau=10^{12}~\text{s}^{-1}$ the relative hight of the resonance
will be $\sim 10^{-3}$.

 Let us emphasize again
that the advantage of a set-up with a polar dielectric coating lies
in the fact that the resonance (\ref{last}) is {\it purely} due to
the intersubband transitions with no other excitations existing in
the same frequency domain that could otherwise obscure their
observation.

\section{Summary and conclusions}

Noble metals, such as Au or Bi/Ag alloys, possess the unique property that a substantial density of two-dimensional surface (Tamm-Shockley) states is present when the surface is grown in a particular direction (111) of a cubic lattice. This allows the study of the interplay of 2D and bulk 3D electron liquids. Of particular significance is the fact that the surface confining potential breaks spatial inversion and leads to an {\it intrinsic}, i.e. independent of any disorder potential, spin-orbit interaction. The latter results in the formation of spin-split surface subbands with the splitting reaching a quite significant magnitude of $\Delta \sim 0.1-0.2~ \text{eV}$ at the Fermi level.

The response functions of the surface states have a resonant character for frequencies close to the energy of spin-orbit splitting. This resonance allows optical detection of 2D states, which are otherwise obscured  by a much large number of bulk electrons. In thin films the optical response can be expected to be influenced by the presence of surface to a larger extent. Measurements of reflection and transmission coefficients in the infrared spectrum should be able to reveal the intersubband resonances.

Another phenomenon, which is predicted to bear the signature of Tamm-Shockley states, is associated with surface plasmons, i.e. collective excitations of electron density. Surface plasmons (as well as symmetric plasmons in metal films) are gapless at long wavelengths. This ensures that the plasmon energy intersects the energy of spin intersubband resonances at some wavelength. The electric field is then strongly enhanced by the motion of 2D electrons which results in significant and detectable modifications of the plasmon spectrum.

\acknowledgments

We acknowledge fruitful discussions with J. Gerton, V. Podolsky, M.
Raikh, T. Shahbazyan, A. Shytov, and O. Starykh. The work was
supported by DOE, Award No.~DE-FG02-06ER46313.

\appendix

\section{Solution of kinetic equation}

To solve kinetic equation (\ref{kinetic}) written in the frequency
representation as
\begin{equation}
\label{kinetic_a} \omega \hat{f}_{\bf p}-\frac{1}{2}\Delta_p  [\hat
\eta_{\bf n} ,\hat{f}_{{\bf p}} ]= i\hat{\cal K}_{\bf p},
\end{equation}
we first calculate its commutator with $\hat\eta_{\bf n}$,
\begin{equation}
\label{kinetic_b} \omega [\hat{f}_{\bf p},\hat\eta_{\bf n}]+\Delta_p
 (\hat{f}_{\bf p}-\hat \eta_{\bf n}\hat{f}_{\bf p}\hat \eta_{\bf
n})= i[\hat{\cal K}_{\bf p},\eta_{\bf n}],
\end{equation}
where we utilized that $\hat \eta_{\bf n}^2=1$. Finally, we need one
more equation, which is found from Eq.~(\ref{kinetic_a})
\begin{equation}
\label{kinetic_c} \omega \hat \eta_{\bf n}\hat{f}_{\bf p}\hat
\eta_{\bf n}+\frac{1}{2}\Delta_p  [\hat \eta_{\bf n} ,\hat{f}_{{\bf
p}} ]= i\hat \eta_{\bf n}\hat{\cal K}_{\bf p}\hat \eta_{\bf n},
\end{equation}
Formally, the system of three equations
(\ref{kinetic_a}-\ref{kinetic_c}) contains three unknowns:
$\hat{f}_{\bf p},$ $[\hat{f}_{\bf p},\hat\eta_{\bf n}]$, and $\hat
\eta_{\bf n}\hat{f}_{\bf p}\hat \eta_{\bf n}$. Eliminating the last
two of these unknowns we obtain the solution of kinetic equation in
the form, Eq.~(\ref{anzats}).

\section{Boundary conditions}

Let us assume that the amplitude of the electromagnetic field acting
on surface electrons is small compared with the atomic fields (this
is a rather weak condition). The modification of the electronic wave
function  can then be neglected and the effective electric field
$E_x$ in Eq.~(\ref{kinetic}) acting on surface electrons is written
as
\begin{equation}
\label{effe} E_x =\int dz |\psi(z)|^2 E_x(z),
\end{equation}
where $\psi(z)$ is the wave function describing transverse
confinement of surface states. Here we adopt the simplest
approximation that $\psi(z)$ is momentum-independent and constant
across the interface layer of thickness $a$. Equation (\ref{effe})
then gives,
\begin{equation}
\label{effecti} E_x =\frac{1}{a} \int_0^a dz E_x(z).
\end{equation}
The interface layer can now be considered as a metal film with the
effective dielectric function
\begin{equation}
\widetilde \varepsilon =-\frac{4\pi e^2
n_2}{m^*a\omega^2}-\frac{m^*}{2\omega^2p_F^2 a} \frac{e^2
\Delta^4}{(\omega+i/\tau)^2-\Delta^2},
\end{equation}
that is obtained from Eq.~(\ref{current}): $n_2$ is 2D density of
surface electrons. Solution of Maxwell's equation for $0<z<a$, which
satisfy the condition that $E(0)=E_1$ and $E(a)=E_2$, is
\begin{equation}
E_x(z) =E_1 \cosh{\kappa z}+(E_2-E_1\cosh{\kappa
a})\frac{\sinh{\kappa z}}{\sinh{\kappa a}},
\end{equation}
where $\kappa=\sqrt{k^2-\widetilde\varepsilon \omega^2/c^2}$. As a
result, Eq.~(\ref{effecti}) gives the effective field,
\begin{equation}
\label{effectiff} E_x =E_1 \frac{\sinh{\kappa a}}{\kappa
a}+(E_2-E_1\cosh{\kappa a})\frac{\cosh{\kappa a}-1}{\kappa a
\sinh{\kappa a}}.
\end{equation}
This condition reduces to
\begin{equation}
\label{effectiff} E_x =\frac{E_1+E_2}{2},
\end{equation}
used throughout the paper as long as $\kappa a \ll 1$. This
condition is satisfied as long as
\begin{equation}
\text{max}(|\omega-\Delta|,1/\tau) \gg \frac{m^*e^2\Delta^3 a}{p_F^2
c^2} \sim 10^{-3} \text{meV}.
\end{equation}

\end{document}